\documentclass[preprint,superscriptaddress,floatfix,prd]{revtex4}
\usepackage{graphicx,xcolor,soul}

\begin{document}

\title{Maximal Subvacuum Effects: A Single Mode Example}

\author{Anastasia Korolov}
\email{akorolov@umd.edu}
\affiliation{ Institute for Research in Electronics and Applied Physics \\
 University of Maryland, College Park, Maryland 20742, USA}

\author{L. H. Ford}
\email{ford@cosmos.phy.tufts.edu}
\affiliation{Institute of Cosmology, Department of Physics and Astronomy \\
Tufts University, Medford, Massachusetts 02155, USA}

\begin{abstract}
We discuss an example of a subvacuum effect, where a quantum expectation value is below the vacuum level, and is hence negative. 
The example is the time average of the mean squared electric field in a non-classical state where one mode is excited. 
We give some specific examples of 
such states, and discuss the lower bound on the squared field or its time average. We show when a lower bound can be obtained by 
diagonalization of the squared electric field operator, and calculate this bound. We also discuss the case of an instant time
mean squared electric field,  when the operator cannot be 
diagonalized. In this case, a lower bound still exists but is attained only by  the limit of a sequence of quantum states. In general, the 
optimum lower bound on the mean squared electric field  is minus one-half of the  mean squared electric field in a one photon state.
This provides a convenient estimate of the subvacuum effect, and may be useful for attempts to experimentally measure this effect.
\end{abstract}

\maketitle
\baselineskip=18pt	

\section{Introduction}
\label{sec:intro}

It is well known that quantities such as the energy density or squared fields, which are positive in classical physics, can acquire negative expectation
values in quantum field theory. It was proven by Epstein {\it et al}~\cite{EGJ65} that this is a general feature of all quantum field theories. Reviews of negative
energy density and its effects are given in Refs.~\cite{roman04,F09}.
Examples include the electromagnetic energy density in the Casimir effect, or in nonclassical quantum states. Effects
such as negative energy density can arise because we are dealing with a renormalized expectation value, from which a formally infinite quantity has been
subtracted. In boundary-free flat spacetime, this means normal ordering of the relevant operator. The expectation value of the normal-ordered operator 
vanishes in the vacuum state, and is positive in states describing classical excitations. Thus when the mean energy density or squared electric field
becomes negative, it is below the vacuum level, and we refer to this as a subvacuum effect. There has been considerable interest in recent years
in the phenomenon of negative energy density, or violation of the weak energy condition, because of its potential gravitational effects. However, the
magnitude and duration of  negative energy density and other subvacuum effects is constrained by quantum 
inequalities~\cite{ford78,ford91,fr95,fr97,flanagan97,FE98,pfenning01,fh05}. For massless fields in four dimensional spacetime, these are inequalities of the
form
\begin{equation}
\int_{-\infty}^\infty dt\, g(t)\, \rho(t) \geq -\frac{C}{\tau^4} \,.
\label{eq:QI}
\end{equation}
Here $\rho(t)$ is the expectation value of the energy density or other classically positive operator in an arbitrary quantum state at a given spatial point,
$C$ is a positive constant, and $g(t)$ is a sampling function in time with characteristic width $\tau$. We normalize the sampling function by
\begin{equation}
\int_{-\infty}^\infty g(t) \, dt =1 \,.
\label{eq:g-norm}
\end{equation}
In the $\hbar = c =1$ units, which we adopt in this paper, $C$ is dimensionless, and typically smaller than unity.

The quantum inequalities severely constrain large subvacuum effects, but this does not mean that such effects are unobservable. Several proposals have been
made for systems where these effects might be large enough to observe. These include transient increases in the magnetization of a spin system~\cite{fgo92},
increases in the lifetimes of excited atoms in a cavity~\cite{FR11}, and increases in the speeds of pulses in a nonlinear material~\cite{DF18}. The latter two effects
involve negative mean squared electric field, which will be of special interest in the present paper. The role of the squared electric field operator in a nonlinear
material as an analog for the effects of negative energy density and its fluctuations in gravity theory was discussed in Refs.~\cite{BDF14,BDFR16}.

With the exception of the two-spacetime dimensional inequality of Ref.~\cite{flanagan97}, the quantum inequalities are not known to be optimal. That is, it is not known
whether the lower bound is actually attained by any quantum state. The optimal lower bound in Eq.~(\ref{eq:QI}) would be the lowest eigenvalue of the averaged operator.
This is the usual situation in quantum mechanics, where the smallest value which may be found in any measurement of an observable is the lowest eigenvalue
of the associated operator. The lowest eigenvalue is also the lowest bound on the expectation value of the observable in any quantum state, and a bound which is
only attained when the quantum state is the eigenstate associated with the lowest eigenvalue.
Thus finding this lowest eigenvalue is one way to compute optimal bounds. A method for diagonalizing quadratic operators was given by Colpa~\cite{Colpa1978}.
A numerical implementation was developed by Dawson~\cite{Dawson2006}, and was  recently used in Ref.~\cite{SFF18} in a study of the probability distribution for 
stress tensor fluctuations. An alternative method for estimating the optimum lower bound from the moments of the operator was discussed in Ref.~\cite{FFR12}.
In the present paper, which is based in part on Ref.~\cite{K17},
we will be concerned with the case of the squared electric field when one mode is excited. This is the case which
most relevant to the experiments proposed in Refs.~\cite{FR11,DF18}, and one where the diagonalization may be performed in closed form. 

In Sect.~\ref{sec:sub}, we give some explicit examples of quantum states leading to a subvacuum effect, a negative mean squared electric field. The squared electric
field operator for a single mode is diagonalized in Sect~\ref{sec:diag}, and the state which minimizes the expectation value of this operator is discussed. A related discussion
of this diagonalization was recently given in Ref.~\cite{SFF18}. The lowest eigenvalue, which gives the maximal subvacuum effect is also obtained in Sect~\ref{sec:diag}.
The physical meaning of the associated lower bound is further discussed in Sect~\ref{sec:meaning}, and the results of the paper are summarized in Sect~\ref{sec:sum}. 
Throughout this paper we use Lorentz-Heaviside units with $\hbar=c=1$.

\section{Subvacuum Effects}
\label{sec:sub}

In this section, we will illustrate the concept of a subvacuum effect, a negative expectation value of a classically positive quantity, with some explicit examples. 
Here we consider the case of the squared electric field for a single mode. The electric field operator in this case may
be written as
\begin{equation}
\mathbf{E}(\mathbf{x},t) = a \, \mathbf{f}(\mathbf{x}) \, {\rm e}^{-i \omega t} +\, a^\dagger\, \mathbf{f}^*(\mathbf{x})  \, {\rm e}^{i \omega t}\,,
\end{equation}
where $a$ and $a^\dagger$ are the annihilation and creation operators for a mode with spatial mode function $ \mathbf{f}(\mathbf{x})$ and angular frequency
$\omega$. This could be a standing mode in a resonant cavity, for example. The normal-ordered squared field is
\begin{equation}
:E^2(\mathbf{x},t): = 2 | \mathbf{f}(\mathbf{x})|^2\, a^\dagger a +  a^2 \,  \mathbf{f}(\mathbf{x})\cdot  \mathbf{f}(\mathbf{x})  \, {\rm e}^{- 2i \omega t}\,+ (a^\dagger)^2 \, 
 \mathbf{f^*}(\mathbf{x})\cdot  \mathbf{f^*}(\mathbf{x}) \, {\rm e}^{2i \omega t}\,.
 \label{eq:E2}
\end{equation}

In some situations, we may be more interested in a time or space average of $:E^2(\mathbf{x},t):$ over a finite region. This is a better model of the response of
a physical detector which requires a finite time to perform a measurement. Here we consider only a time average, which may be described by a sampling function
$g(t)$, which we take to be real, even, and normalized by Eq.~(\ref{eq:g-norm}).
The time average of the squared electric field at a fixed spatial point becomes
\begin{equation}
T = \int_{-\infty}^\infty :E^2(\mathbf{x},t): \, g(t) \, dt = 
 2 | \mathbf{f}(\mathbf{x})|^2\, a^\dagger a +  \left[ a^2 \,  \mathbf{f}(\mathbf{x})\cdot  \mathbf{f}(\mathbf{x}) + (a^\dagger)^2 \, \mathbf{f^*}(\mathbf{x})\cdot  \mathbf{f^*}(\mathbf{x}) \right]
 \, \hat{g}(2 \omega) \,,
 \label{eq:T0}
\end{equation}
where
\begin{equation}
\hat{g}( \omega) = \int_{-\infty}^\infty g(t) \,  {\rm e}^{i \omega t}\, dt 
\end{equation}
is the Fourier transform of $g(t)$, and is itself real and even. We can write $T$ as
\begin{equation}
T = A \,a^{\dagger} a + B\, a^{2} + B^* \, (a^\dagger)^2 \,,
\label{eq:T1}
\end{equation}
where
\begin{equation}
A =  2 | \mathbf{f}(\mathbf{x})|^2 > 0\,,
\label{eq:A}
\end{equation}
and
\begin{equation}
B = \mathbf{f}(\mathbf{x})\cdot  \mathbf{f}(\mathbf{x})\, \hat{g}(2 \omega) \,.
\label{eq:B}
\end{equation}
The local energy density operator or that for the time-averaged energy density for a single mode may also be written in the form of Eq.~(\ref{eq:T1}).

Now we wish to illustrate how subvacuum effects for the squared electric field or the energy density can arise in certain quantum states. One simple example is a
superposition of the vacuum and a two-particle state
\begin{equation}
|\psi\rangle = \frac{1}{\sqrt{1+ \epsilon^{2}}}(|0\rangle + \epsilon |2\rangle)\,,
\label{eq:0+2}
\end{equation}
where we may take $\epsilon$ to be real. The expectation value of $T$ in this state becomes
\begin{equation}
\langle T \rangle =  \frac{\epsilon}{1+ \epsilon^{2}} \, \left[ \sqrt{2} (B + B^*) + 4 \epsilon \, A \right]\,
\end{equation}
which will be negative if $\epsilon$ is chosen to have the opposite sign as $B + B^*$ and $|\epsilon| < \sqrt{2} |B + B^*|/(4 A)$. In this example, the subvacuum
effect of negative squared electric field or energy density arises as a quantum interference effect between states of different particle number. 
Note that the mean number of particles in the state $|\psi\rangle$ is
\begin{equation}
\langle n \rangle =  \frac{2 \epsilon^2}{1+ \epsilon^{2}} \, ,
\end{equation}
and hence will be small if $|\epsilon| \ll 1$.

Another class of quantum states which lead to subvacuum effects are the squeezed vacuum states, introduced by Stoller~\cite{Stoller}, and reviewed
in Refs.~\cite{Caves,GN}. 
These are a one complex parameter family of states defined by
\begin{equation}
|\zeta\rangle = S(\zeta) \, |0\rangle\,,
\label{eq:sq-vac}
\end{equation}
where the squeeze operator is defined by
\begin{equation}
S(\zeta) = \exp\left[ \frac{1}{2} (\zeta^{*} a^{2} - \zeta a^{\dagger 2}) \right] \,.
\label{eq:sq-op}
\end{equation}
This is a unitary operator which satisfies $S(-\zeta) = S^\dagger(\zeta) = S^{-1}(\zeta)$. 
The squeezed vacuum states may arise from quantum particle creation effects, as described for example in Sect. 7.2 of Ref.~\cite{GN}. When a classical pump field passes through a
nonlinear material with nonzero third-order susceptibility, the squeezed vacuum is created by degenerate four-wave mixing as was first done by 
Slusher, {\it et al}~\cite{Slusher}. If a material with a nonzero second-order susceptibility is used, the squeezed vacuum is generated by degenerate parametric down-conversion,
as was first achieved by Wu, {\it et al}~\cite{Wu}.

We can see from Eqs.~(\ref{eq:sq-vac}) and (\ref{eq:sq-op}) that the squeezed vacuum state, $|\zeta\rangle $, is a superposition of all possible even number particle eigenstates.
It may be shown~\cite{Caves,GN} that
\begin{equation}
S^{\dagger} a S = a \cosh{r} - a^{\dagger} {\rm e}^{i \delta} \sinh{r} \,,
\label{eq:SdaS}
\end{equation}
where $\zeta = r \, {\rm e}^{i \delta}$. From this relation, it follows that
\begin{equation}
\langle\zeta| a^{2} |\zeta\rangle= \langle\zeta|  a^{\dagger 2} |\zeta\rangle^{*} = -{\rm e}^{i \delta} \sinh{r} \cosh{r} 
\end{equation}
and that the mean number of particles in the state $|\zeta\rangle $ is
\begin{equation}
\langle n \rangle =   \langle\zeta| a^{\dagger}\, a |\zeta\rangle = \sinh^{2}{r}\,.
\label{eq:n-sqvac}
\end{equation}

If we take the expectation value of Eq.~(\ref{eq:T1}) in this state, the result may be written as
\begin{equation}
\langle T \rangle = \sinh r \, [ A\, \sinh r - 2 \cosh r \, {\rm Re}(B\, {\rm e}^{i \delta})]\,.
\end{equation}
Consider the case where $T$ is the squared electric field given in Eq.~(\ref{eq:E2}), and assume that $\delta =0$ and that the mode function $ \mathbf{f}$ is real.
Then we have
\begin{equation}
A  =  2\,  \mathbf{f}^2(\mathbf{x}) \qquad {\rm and } \qquad B  =   \mathbf{f}^2(\mathbf{x})\,   {\rm e}^{- 2i \omega t}\,,
\label{eq:ABE2}
\end{equation}
so
\begin{equation}
\langle  :E^2(\mathbf{x},t): \rangle =  2  \mathbf{f}^2(\mathbf{x})\, \sinh r \, [\sinh r - \cosh r \, \cos(2\, \omega \,t) ]\,.
\label{eq:<E2>}
\end{equation}
Note that $\langle  :E^2(\mathbf{x},t): \rangle $  is a periodic function of time at a fixed spatial point.  Because $\cosh r > \sinh r$,  it will be negative near $t=0$.
However, its time average over a sufficiently long time will be positive. Note that when $r \ll 1$, the squeezed vacuum state $|\zeta\rangle $ is approximately 
the vacuum plus two-particle state of Eq.~(\ref{eq:0+2}) with $\epsilon \approx r/\sqrt{2}$.

\section{Diagonalization}
\label{sec:diag}

In this section, we treat the diagonalization of quadratic operators for a single mode, such as those describing energy density or squared electric field.
The basic strategy will be to change the basis of creation and annihilation operators in such a way as to find the eigenstates and eigenvalues of the
original operator.

Consider an operator of the form of Eq,~(\ref{eq:T1}), where $a$ and $a^{\dagger} $ are the annihilation and creation operators for a single mode of any 
bosonic quantum field, $A>0$ is a real constant, and $B$ is a complex constant. We wish to find a Bogolubov transformation of the form
\begin{equation}
a = \alpha b + \beta b^{\dagger}\,,
\label{eq:B-tranform}
\end{equation}
where $b$ and $b^{\dagger}$ form a pair of annihilation and creation operators for the same mode. The commutation relations, $[a,a^{\dagger}] = [b,b^{\dagger}]=1$,
require that the constants $\alpha$ and $\beta$ satisfy
 \begin{equation}
|\alpha|^{2} - |\beta|^{2} = 1 \,.
\label{eq:norm}
\end{equation} 
If we substitute  Eq.~(\ref{eq:B-tranform}) into Eq.~(\ref{eq:T1}), the result may be expressed as
\begin{eqnarray}
T &=& [A(|\alpha|^{2} +  |\beta|^{2} )  + 2B\alpha\beta  +  2 B^* \alpha^* \beta^* ]\, b^{\dagger}b 
+ [ A\alpha\beta^* + B\alpha^{2} + B^* (\beta^*)^{2}] \, b^2 \nonumber \\
&+& [ A\alpha^* \beta + B^* (\alpha^*)^{2} + B \beta^{2}] \, (b^\dagger)^2 
+ A\, |\beta|^{2} + B\alpha\beta + B^* \alpha^* \beta^* \,.
\label{eq:T2}
\end{eqnarray}

We now wish to impose the diagonalization condition,
 \begin{equation}
 A\alpha\beta^* + B\alpha^{2} + B^* (\beta^*)^{2} =0 \,,
 \label{eq:diag1}
 \end{equation} 
to remove the $b^2$ and $(b^\dagger)^2$ terms in $T$. Write $B = |B|\, {\rm e}^{i \gamma}$, $\alpha = |\alpha|\, {\rm e}^{i \eta}$, and $\beta = |\beta|\, {\rm e}^{i \delta}$, 
to write this condition as 
\begin{equation}
A|\alpha||\beta| + |B| \left(|\alpha|^{2} \,  {\rm e}^{i (\gamma+\eta +\delta)}   +  |\beta|^{2} \,  {\rm e}^{-i (\gamma+\eta +\delta)} \right) =0 \,
 \label{eq:diag2}
 \end{equation} 
However, because $A>0$ and $|\alpha| > |\beta|$, this condition may only be satisfied if ${\rm e}^{i (\gamma+\eta +\delta)} =-1$, which is fulfilled if
\begin{equation}
\gamma+\eta +\delta = \pi\,,
\end{equation}
We take a solution of this condition where $\eta = 0$, so $\alpha$ is real and positive, and where
\begin{equation}
\delta = \pi - \gamma\,.
\end{equation}
so Eq.~(\ref{eq:norm}) becomes
\begin{equation}
\alpha = \sqrt{1+|\beta|^{2}}\,.
\end{equation}
We can now write Eq.~(\ref{eq:diag2}) as an equation for $|\beta|^{2}$, as
\begin{equation}
A^2\, |\beta|^{2} (1 +| \beta|^{2}) = |B|^2 \, (1 + 2 |\beta|^{2})^2\,,
\end{equation}
which has the solution
\begin{equation}
|\beta| =  \sqrt{- \frac{1}{2} + \frac{1}{2}\sqrt{\frac{A^{2}}{A^{2}-4|B|^{2}}}}
\label{eq:beta}
\end{equation}
leading to
\begin{equation}
\alpha =  \sqrt{\frac{1}{2} + \frac{1}{2}\sqrt{\frac{A^{2}}{A^{2}-4|B|^{2}}}} \,.
\label{eq:alpha}
\end{equation}
Note that this solution is meaningful only if
\begin{equation}
A > 2 |B|\, ,
\label{eq:diag-condition}
\end{equation}
which is the condition that $T$ be diagonalizable.

If we substitute Eqs,~(\ref{eq:beta}) and (\ref{eq:alpha})  into Eq.~(\ref{eq:T1}), and note that $B \beta = B^* \beta^* = - |B|  |\beta|$,
we find the diagonal form of $T$,
\begin{equation}
T = \Omega \, b^{\dagger}b  + \lambda_0\,.
\label{eq:T3}
\end{equation}
Here 
\begin{equation}
\Omega = A(|\alpha|^{2} +  |\beta|^{2} )  + 2B\alpha\beta  +  2 B^* \alpha^* \beta^* = \sqrt{A^2 -4 |B|^2}\, ,
\end{equation}
and
\begin{equation}
\lambda_0 = A\, |\beta|^{2} + B\alpha\beta + B^* \alpha^* \beta^* = \frac{1}{2}\, \left( \sqrt{A^2 -4 |B|^2} -A \right)\,.
\end{equation}
Note that Eq.~(\ref{eq:T3}) has the form of a quantum harmonic oscillator Hamiltonian with frequency $\Omega$, and zero point energy $\lambda_0$.
However, $T$ has the physical interpretation of a local operator, such as energy density, or the time average over a finite interval of such an operator.
Note that in the limit $B \rightarrow 0$, we obtain $\Omega \rightarrow A$ and $\lambda_0  \rightarrow 0$, which is consistent with Eq.~(\ref{eq:T1}).
  
The eigenstates of $T$ are the number eigenstates  $| n\rangle_b$ in the $b$-basis, 
\begin{equation}
T | n\rangle_b = \lambda_n \, | n\rangle_b \,,
\end{equation}
with eigenvalues  $\lambda_n = n \, \Omega +  \lambda_0$ for $n = 0, 1, 2, \ldots$.
We are especially interested in the lowest eigenvalue, $\lambda_0$, which is associated with the $b$-vacuum state, $| 0\rangle_b$.
First, we note that this eigenvalue is always negative if $B \not= 0$,
 \begin{equation}
\lambda_0 = \frac{1}{2}\, \left( \sqrt{A^2 -4 |B|^2} -A \right) <  0 \,.
\label{eq:lambda0}
\end{equation}  
As the lowest eigenvalue of $T$, it represents the maximal subvacuum effect when a single mode is excited.
Furthermore,
\begin{equation}
\lambda_0 > -  \frac{1}{2}\, A \,,
\label{eq:QI2}
\end{equation}
with this lower bound on $\lambda_0$ being approached arbitrarily closely in the limit $A \rightarrow 2 |B|$.  
Equation~(\ref{eq:QI2}) is a quantum inequality bound on the expectation value of $T$ in any quantum state, and hence
on the magnitude of any subvacuum effect for the observable associated with this operator.

The associated state may be represented in the $a$-Fock space as
\begin{equation}
| 0\rangle_b =  \sum_{n=0}^\infty c_n \,  |n\rangle_a \,.
\label{eq:expansion}
\end{equation}
The inverse transformation to Eq.~(\ref{eq:B-tranform}) is
\begin{equation}
b = \alpha a - \beta a^{\dagger}\,.
\label{eq:inv-B-tranform}
\end{equation}
If we act on Eq.~(\ref{eq:expansion}) with this relation, the result is
\begin{equation}
 \sum_{n=0}^\infty \left[ \alpha c_{n} \sqrt{n} |{n-1}\rangle_{a} - \beta c_{n} \sqrt{n+1} |{n+1}\rangle_{a} \right]  = 0 \,,
\end{equation}
which may be written as
\begin{equation}
 \sum_{n=0}^\infty \left[ \alpha c_{n+1} \sqrt{n+1}  - \beta c_{n-1} \sqrt{n} \right] \, |{n}\rangle_{a }\,.
\end{equation}
This leads to a recurrence relation for the $c_n$,
\begin{equation}
c_{n+1} = \frac{\beta}{\alpha}\sqrt{\frac{n}{n+1}} \; c_{n-1} \,.
\end{equation}
Note that each $c_{n}$ depends only on $c_{n-2}$, so the $c_{n}$ for even $n$ and those for odd $n$  are independent of one another.
We require that $| 0\rangle_b \rightarrow | 0\rangle_a$ as $B \rightarrow 0$, and hence $\beta \rightarrow 0$. This will be achieved if
$c_1 =0$, so $c_{n} =0 $ for odd $n$. The recurrence relation may now be solved to obtain
\begin{equation}
c_{2n} = \sqrt{\frac{(2n-1)!!}{(2 n)!!}} \, \left(\frac{\beta}{\alpha} \right)^n \; c_0
\end{equation}
for $n \geq 1$. The normalization condition for the state is $\sum_{n=0}^\infty |c_{2n}|^2 =1$. This may be combined with the identity
\begin{equation}
1 + \sum_{n=1}^\infty  \frac{(2n-1)!!}{(2 n)!!} \, \left(\frac{|\beta|}{\alpha} \right)^{2n} = \frac{\alpha}{\sqrt{\alpha^2 - |\beta|^2}} = \alpha \, ,
\end{equation}
to write the lowest eigenstate of $T$ as
\begin{equation}
 |{0}\rangle_{b} = \frac{1}{\sqrt{\alpha}} \,\left[  |{0}\rangle_{a } + \sum_{n=1}^\infty  \sqrt{\frac{(2n-1)!!}{(2 n)!!}} \, \left(\frac{\beta}{\alpha} \right)^n \; |{2n}\rangle_{a } \right]\,. 
 \label{eq:eigenstate}
 \end{equation}
 
The state $ |{0}\rangle_{b}$  is a squeezed vacuum state in the $a$-Fock space, as may be shown as follows. Note that if we let $\zeta \rightarrow -\zeta$, then
Eq.~(\ref{eq:SdaS}) becomes
\begin{equation}
S(\zeta) a S^\dagger(\zeta) = a \cosh{r} + a^{\dagger} {\rm e}^{i \delta} \sinh{r} \,,
\label{eq:SaSd}
\end{equation}
which is Eq.~(\ref{eq:inv-B-tranform}) with $\alpha = \cosh{r}$ and $\beta = - {\rm e}^{i \delta} \sinh{r}$. Thus, if $b = S(\zeta) a S^\dagger(\zeta)$, then 
$b  |{0}\rangle_{b} = S(\zeta) a S^\dagger(\zeta) S(\zeta) |{0}\rangle_{a} = S(\zeta) a |{0}\rangle_{a} = 0$. This shows that the operator $b$ which annihilates the state
$ |{0}\rangle_{b}$  is also the operator given by the inverse Bogolubov transformation, Eq.~(\ref{eq:inv-B-tranform}), so $ |{0}\rangle_{b} =  S(\zeta) |{0}\rangle_{a}$ 
is a squeezed vacuum state.

The mean number of particles in the lowest eigenstate is
\begin{equation}
\langle n \rangle_0 =  {}_b \langle 0| a^\dagger a \,  |{0}\rangle_{b} = 
{}_b \langle 0|  ( \alpha b^{\dagger} + \beta^* b)  ( \alpha b + \beta b^{\dagger}) \,  |{0}\rangle_{b} = |\beta|^2\,,
\end{equation}
which in our case becomes
\begin{equation}
\langle n \rangle_0 = \frac{A}{2 \, \sqrt{A^2 -4 |B|^2}} - \frac{1}{2} \,.
\label{eq:mean-num}
\end{equation}
Note that $\langle n \rangle_0 \rightarrow \infty$ as $|B| \rightarrow \frac{1}{2} A$, the limit in which $T$ ceases to be diagonalizable.
However, unless $2 |B|/A$ is very close to one, $\langle n \rangle_0$ is of order one. For example, $2 |B|/A = 0.98$ leads to  
$\langle n \rangle_0 \approx 2.0$. Thus, in many cases, a vacuum plus two particle state, Eq.~(\ref{eq:0+2}), is a fair approximation
to the lowest eigenstate,  $|{0}\rangle_{b}$, of $T$. 

It is instructive to return to the instant
time  expectation value of the local squared electric field in a general squeezed vacuum state, Eq.~(\ref{eq:<E2>}). Note, from
Eq.~(\ref{eq:ABE2}), that $A = 2 |B|$ in this case, so the operator is not diagonalizable. The maximally negative expectation value occurs
at $t=0$, where
\begin{equation}
\langle  :E^2(\mathbf{x},0): \rangle =  2  \mathbf{f}^2(\mathbf{x}) \, \sinh r \, (\sinh r - \cosh r )
   =    \mathbf{f}(\mathbf{x})^2\, (-1 + {\rm e}^{-2r} ) \,.
\label{eq:<E2>b}
\end{equation}
We see that this value approaches a lower bound of $-  \mathbf{f}^2(\mathbf{x})$, but only in the limit that $r \rightarrow \infty$, corresponding
to an infinite mean number of particles. For the operator to be diagonalizable, there would have to exist a state corresponding to the lowest
eigenvalue and hence lowest expectation value. In this case, the lowest expectation value is only approached as the limit of an infinite sequence
of states.
However, it is only the squared electric field at one time which fails to be diagonalizable. Any time averaging, as in Eq.~(\ref{eq:T0}),
 will result in a diagonalizable operator. We may see this from Eqs.~(\ref{eq:A}) and (\ref{eq:B}) and from
\begin{equation}
\hat{g}(2 \omega) =  \int_{-\infty}^\infty g(t) \,  \cos( 2 \omega t) \, dt <  \int_{-\infty}^\infty g(t) \, dt  =1 \,.
\end{equation}

\begin{figure}[htbp]
	\centering
		\includegraphics[scale=0.4]{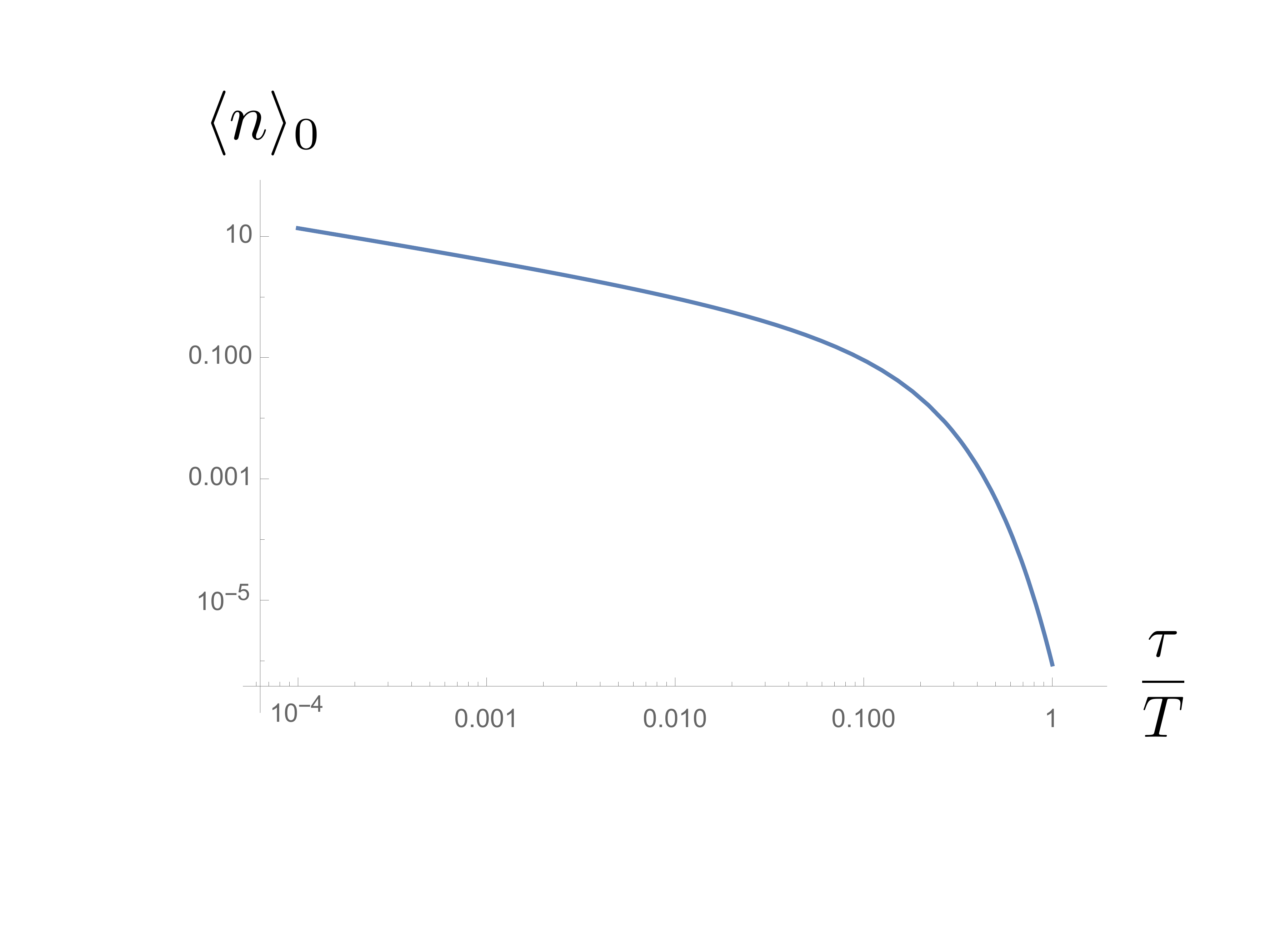}
		\caption{The mean number of particles in the lowest eigenstate of the Lorentzian averaged squared
		electric field is plotted as a ratio of the sampling time $\tau$ and the period $T$ of the excited mode.
		This is the state which maximizes the subvacuum effect in this case, and we see that unless $\tau$
		is very small, this mean number is relatively small.}
	\label{fig:n-plot}
\end{figure}

Thus the time-averaged squared electric field has a well-defined lowest eigenstate, which represents the maximal subvacuum effect. The mean 
number of photons in this state is given by Eqs.~(\ref{eq:A}), (\ref{eq:B}) and  (\ref{eq:mean-num}) to be
\begin{equation}
\langle n \rangle_0 = \frac{1}{2 \, \sqrt{1 - \hat{g}(2 \omega) } }- \frac{1}{2} \,.
\label{eq:mean-num-2}
\end{equation}
Consider the explicit case of a Lorenztian sampling function of width $\tau$,
\begin{equation}
g_L(t) = \frac{\tau}{\pi(t^2 +\tau^2)}\,,
\end{equation}
for which
\begin{equation}
\hat{g_L}(2 \omega) =  \int_{-\infty}^\infty g(t) \, \cos( 2\omega t) \, dt = {\rm e}^{-2 \omega \tau}\,.
\label{eq:g_L}
\end{equation}
In this case, the mean number of particles in the lowest eigenstate is 
\begin{equation}
\langle n \rangle_0 = \frac{1}{2 \, \sqrt{1 - {\rm e}^{-4 \pi \tau/T} } }- \frac{1}{2} \,.
\end{equation}
where $T= 2\pi/\omega$ is the period of the excited mode. This number is very small unless $\tau \ll T$, as is 
illustrated in Fig.~\ref{fig:n-plot}.

\begin{figure}[htbp]
	\centering
		\includegraphics[scale=0.45]{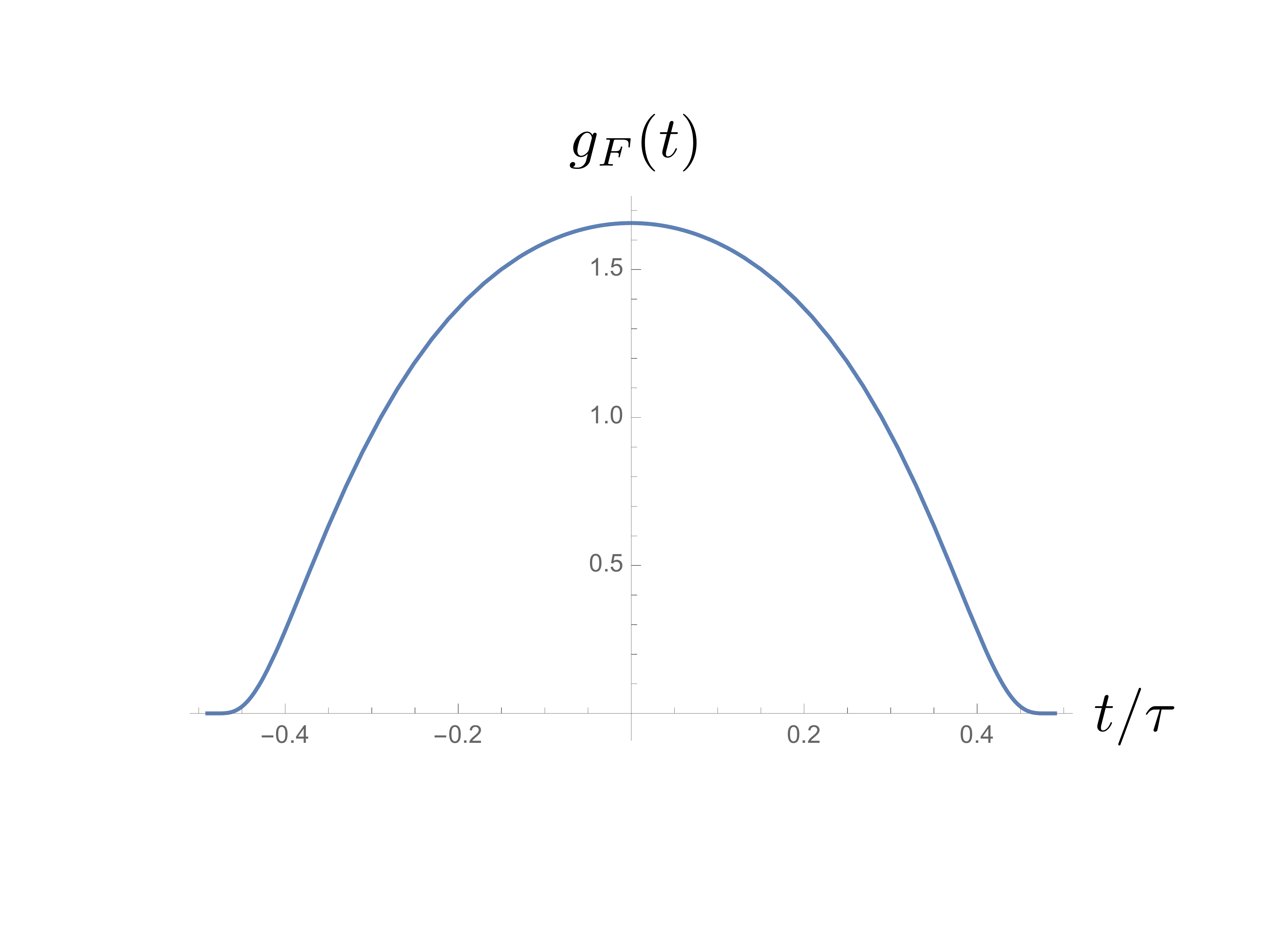}
		\caption{The finite duration function,  $g_F(t)$ is plotted.}
	\label{fig:gF}
\end{figure}

\begin{figure}[htbp]
	\centering
		\includegraphics[scale=0.45]{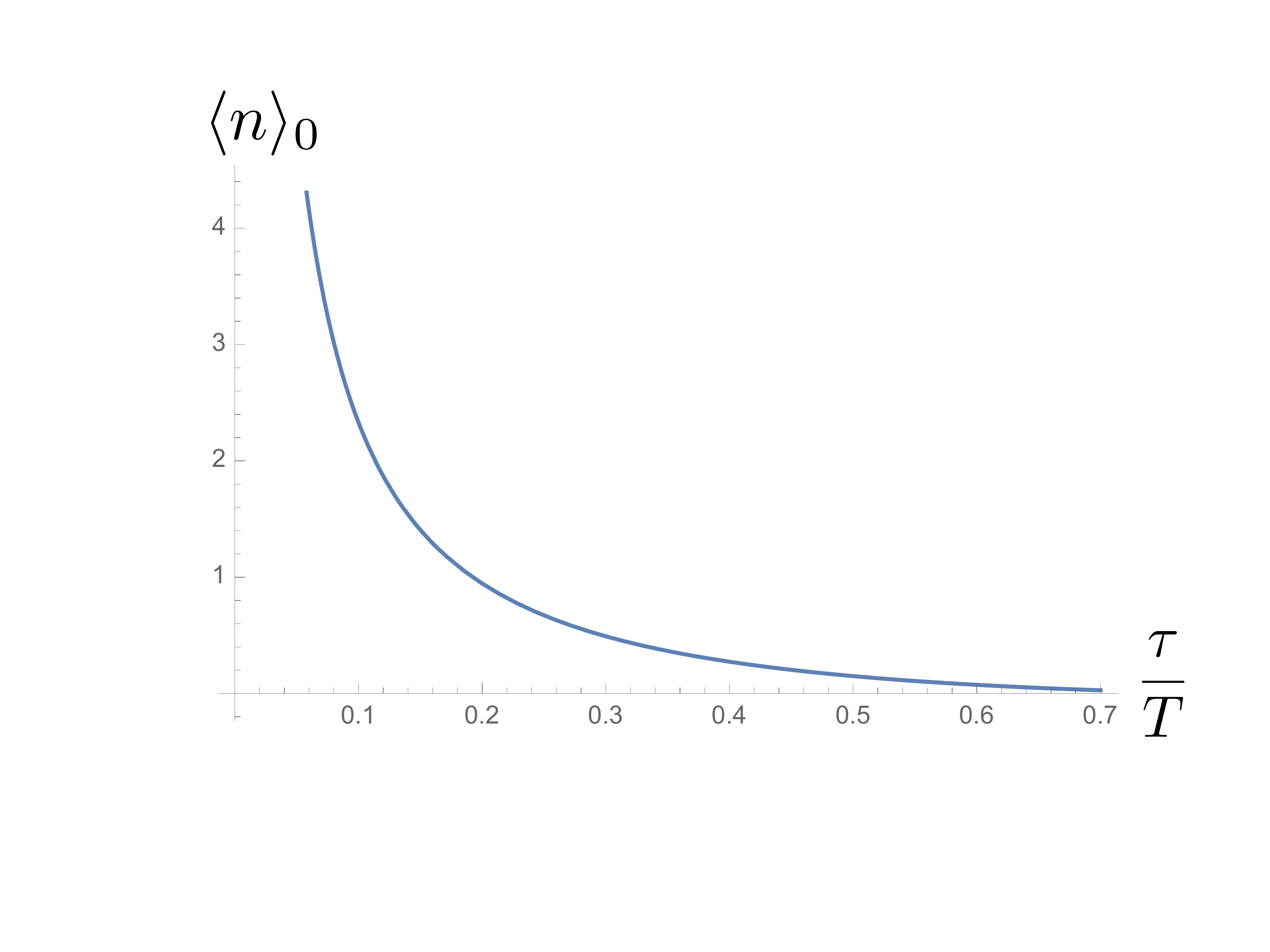}
		\caption{Here the mean number of particles in the lowest eigenstate of the  squared
		electric field averaged over a finite time is plotted as a function of $\tau/T$.  
		 In this case, this mean number is larger than for the case of Lorentzian averaging.}
	\label{fig:n-plot-compact}
\end{figure}

Another class of sampling functions of interest are functions with a finite temporal duration, that is, functions of compact
support. These are more realistic descriptions of a measurement than functions such as Lorentzians or Gaussians, which
have infinite tails into the past. In field theory with an infinite number of degrees of freedom, the probability of large stress
tensor fluctuations is greatly enhanced when the sampling functions have finite duration~\cite{FF15}. An example of such a function
is defined by
\begin{equation}
g_F(t) = \frac{K}{\tau}\, \exp\left[-\frac{\tau^2}{4 ( \frac{1}{2} \tau +t)  ( \frac{1}{2} \tau  -t)}\right]\,, \quad - \frac{1}{2} \tau < t <  \frac{1}{2} \tau \,,
\end{equation}
and $g_F(t) =0$ if $|t| \geq \frac{1}{2} \tau$.  Here $K \approx 4.50457$, and is found from numerical integration of Eq.~(\ref{eq:g-norm}).
 This non-analytic, but infinitely differentiable function has duration
$\tau$ and is plotted in Fig.~\ref{fig:gF}
As was shown in Ref.~\cite{FF15}, the switch-on part of this function near $\tau = - \frac{1}{2} \tau$ accurately describes the rise in
current in a simple electrical circuit just after the switch is closed. Thus functions of this type may be regarded as reasonable models
for physical switching processes. 
The Fourier transform, $ \hat{g_F}(2 \omega)$, may be computed numerically and used to find $ \langle n \rangle_0$ for this case,
which is plotted in Fig.~\ref{fig:n-plot-compact}.
Note that the mean number of photons in the lowest eigenstate tend to be larger for given $\tau/T$. in the case of $g_F(t)$, as compared to
$g_L(t)$, but is still of order one unless $\tau \ll T$.

\section{Physical Meaning of the Lower Bound}
\label{sec:meaning}

 In this section, we will examine the physical interpretation of the results obtained in the previous section. The key result is the lowest eigenvalue $\lambda_0$
 of the operator $T$, given in Eq.~(\ref{eq:lambda0}). We examine the case where $T$ is the time averaged squared electric field given in Eq.~(\ref{eq:T0}) and 
 the spatial mode function $ \mathbf{f}(\mathbf{x})$ is real. In this case,
 \begin{equation}
A =  2 \,  \mathbf{f}^2(\mathbf{x}) \,,
\label{eq:A1}
\end{equation}
and
\begin{equation}
B = \frac{1}{2} A\, \hat{g}(2 \omega) \,,
\label{eq:B1}
\end{equation}
 and the lowest eigenvalue becomes
 \begin{equation}
\lambda_0 = -   \mathbf{f}^2(\mathbf{x})\, \left(1 -\sqrt{1 - \hat{g}^2(2 \omega)} \right)\,. 
\end{equation}
This is the smallest value which could be found in an individual measurement of the time averaged squared electric field. 

Consider the case where the measurement occurs on a time scale which is small compared to the temporal period of oscillation
of the mode, so that $ \hat{g}(2 \omega) \approx 1$, (See Eq.~(\ref{eq:g_L}), for example.) and hence
\begin{equation}
\lambda_0 \approx -  \mathbf{f}^2(\mathbf{x}) \,.
\label{eq:maxsubvac}
\end{equation}
We can gain some insight into the meaning of this result by noting that the expectation value of T in a one-photon state
is
\begin{equation}
\langle 1|T|1 \rangle = A = 2 \,  \mathbf{f}^2(\mathbf{x}) \,.
\end{equation}
This means that the maximal subvacuum effect involves a decrease below the vacuum level whose magnitude is 
 one-half of the increase when a single photon is added to a cavity in the vacuum state.
In this sense, the maximal subvacuum effect is that of $-1/2$ of a photon. If the effect of adding one photon is observable, then
there is a reasonable chance that the subvacuum effect could also be observable. The subvacuum effect is a suppression
of quantum fluctuations below the vacuum level. Just as the zero point energy of a quantum harmonic oscillator is one-half
of the energy difference between higher energy levels, the maximal subvacuum effect in  a quantity such as squared electric
field is one-half of the effect of adding one photon.
Note that the maximal subvacuum effect described by Eq.~(\ref{eq:maxsubvac}) requires that the cavity be prepared in the special 
quantum state $ |{0}\rangle_{b} $ given in Eq.~(\ref{eq:eigenstate}), and the effect be measured on a sufficiently short time scale.
When this time becomes of the order of or longer than the oscillation period, then both $ \hat{g}(2 \omega)$ and hence the
magnitude of $\lambda_0$ decrease.
 
We can illustrate  the maximal subvacuum effect given by Eq.~(\ref{eq:maxsubvac}) explicitly in the context of the model given
in Ref.~\cite{FR11}.  The key idea of this model is that a squared electric field can alter the decay rate of an atom in an excited
state. An increase in the squared electric field compared to the vacuum state increases the decay rate, and can be viewed as the
effect of stimulated emission. In contrast, a decrease in the squared electric field below the vacuum level decreases the decay rate.
This can be viewed as a suppression of the vacuum fluctuations which are essential for spontaneous emission. If there were no
coupling of the atom to the quantized electromagnetic field, all of its energy levels could be eigenstates of the Hamiltonian, and
hence stable. In the model of  Ref.~\cite{FR11}, the atom passes through one direction of a rectangular cavity which is small compared
to the other dimensions. The lowest mode of the cavity is a TE mode whose frequency depends upon the longer two dimensions.
At least in principle, the transit time of the atom through the shortest direction could be made smaller than the oscillation period.
Then the decrease in decay rate at  the maximal subvacuum effect is about one-half of the increase which would occur when
one photon is added to a cavity in the vacuum state. 

Another illustration of the implications of Eq.~(\ref{eq:maxsubvac})  comes from the models described in Refs.~\cite{DF18,BDF14}.
Here a probe pulse propagates in a nonlinear material with nonzero third-order susceptibility which also contains photons of longer wavelength
in a squeezed vacuum state. The latter create regions of negative mean squared electric field which in turn increase the speed of
the probe pulse. (Note that the nonlinear material in which this occurs is distinct from the material used to create the squeezed 
state in the first place, and the two materials can be in different locations.)  In this model, it is assumed that dispersion can be neglected
over the frequency bandwidth of the pulse, so that the phase and group velocities are approximately equal.
If the size of the wave packet of the probe pulse is
smaller than the wavelength of the squeezed light, then this wave packet can propagate in a region of nearly maximally negative
squared electric field. In this case, the maximal subvacuum effect described by Eq.~(\ref{eq:maxsubvac}) is nearly attained. It can be estimated
as having a magnitude one-half of the speed decrease produced by adding one photon with the wavelength of the squeezed light to
the vacuum state. 

In both of these models, there is some averaging in space and time produced by the specific experimental configuration. In the case of
the atom in a cavity, the cavity geometry and atom's trajectory define an averaging. In the case of the probe pulse in a nonlinear material,
the geometry of the material and the shape of the pulse wave packet can determine an averaging. In both cases, if the atom or the pulse
can primarily sample the region of  maximally negative squared electric field, the maximal subvacuum effect can occur. Given the form and 
width of the sampling function, calculations of $\langle n \rangle_0$, such as those illustrated in Figs.~\ref{fig:n-plot} and \ref{fig:n-plot-compact},
give us the mean number of photons, and hence squeeze parameter from Eq.~(\ref{eq:n-sqvac}), of the quantum state which leads to the 
maximal effect. Note that this mean number depends upon the form of the sampling function, which is in turn defined by the physical
averaging process.

Finally, we should note that although Eqs.~(\ref{eq:QI}) and (\ref{eq:QI2}) are both quantum inequality bounds, they have very different forms.
The reason for this is that  Eq.~(\ref{eq:QI}) is required to hold for all quantum states in a quantum field theory with an infinite number of degrees
of freedom. This includes states in which modes with arbitrarily short wavelengths are excited. Such modes can produce negative energy
density or negative squared electric field with high magnitudes, but correspondingly short durations, as described by  Eq.~(\ref{eq:QI}). In
contrast, Eq.~(\ref{eq:QI2}) is the bound satisfied by all quantum states where only one mode is excited. In this case, the bound approaches
a finite limit as the sampling time $\tau$ becomes very small.

\section{Summary}
\label{sec:sum}

We have treated negative expectation values of the mean squared electric field operator as an example of a subvacuum effect
which might be observable in an experiment. We consider quantum states in which a single mode of the field is excited, and gave
some examples of states leading to negative expectation values. We then diagonalized the time averaged squared electric field operator,
and constructed its lowest eigenstate, which is a squeezed vacuum state, and the corresponding eigenvalue. This state gives the maximal
subvacuum effect for this operator. The time average is essential for the operator to be diagonalizable. The square of the electric field
at one spacetime point cannot be diagonalized because its lowest expectation value is not achieved by a single quantum state, but
rather is approached asymptotically by a sequence of states. We also calculated the mean number of photons in the lowest eigenstate
of a time-averaged operator, and found that it is of order one if the averaging time is of the order of the period of the mode, but grows
when the averaging time becomes small. The maximal subvacuum effect occurs when the mean squared electric field is negative and
has a magnitude equal to one half of its value in a one photon state for the chosen mode. This provides a convenient estimate of the
magnitude of the subvacuum effect, and suggests that it may be observable in an experiment in which the effect of a single photon
can be measured.

\begin{acknowledgments}
This work was supported in part by the National Science Foundation under Grant PHY-1607118.
\end{acknowledgments}


\begin{thebibliography}{20}
 
 \bibitem{EGJ65} H. Epstein, V. Glaser, and A. Jaffe, {\it Nonpositivity of the energy density in quantized field theories},
  Nuovo Cim. {\bf 36}, 1016 (1965). 
 
 \bibitem{roman04}  T.A. Roman, {\it Some Thoughts on Energy Conditions and Wormholes},
in Proceedings of the Tenth Marcel Grossmann Meeting, 
 (World Scientific, Singapore, 2005), pp 1909-1920,  arXiv:gr-qc/0409090.
 
 \bibitem{F09} L. H. Ford, {\it Negative Energy Densities in Quantum Field Theory},
 Int. J. Mod. Phys. A {\bf25} 2355 (2010), arXiv:0911.3597.

\bibitem{ford78} L. H. Ford, {\it Quantum coherence effects and the second law of thermodynamics}, 
Proc. R. Soc. A {\bf 364}, 227 (1978).

\bibitem{ford91} L. H. Ford, {\it Constraints on negative-energy fluxes},  Phys. Rev. D {\bf 43}, 3972 (1991).

\bibitem{fr95} L. H. Ford and T. A. Roman, 
{\it Averaged Energy Conditions and Quantum Inequalities},  
Phys. Rev. D {\bf 51}, 4277 (1995), arXiv:gr-qc/9410043.

\bibitem{fr97} L. H. Ford and T. A. Roman, {\it Restrictions on Negative Energy Density in Flat Spacetime},  
Phys. Rev.  D {\bf 55}, 2082 (1997),  arXiv:gr-qc/9607003.

\bibitem{flanagan97} E. E. Flanagan, {\it Quantum inequalities in two dimensional Minkowski spacetime},  
Phys. Rev. D {\bf 56}, 4922 (1997),  arXiv:gr-qc/9706006.

\bibitem{FE98} C. J. Fewster and S. P. Eveson, {\it Bounds on negative energy densities in flat spacetime},  
Phys. Rev. D  {\bf 58}, 084010 (1998),
arXiv:gr-qc/9805024. 

\bibitem{pfenning01} M. J. Pfenning, {\it Quantum Inequalities for the Electromagnetic Field},  
Phys. Rev. D {\bf 65}, 024009 (2001), arXiv:gr-qc/0107075.

\bibitem{fh05} C. J. Fewster and S. Hollands, {\it Quantum Energy Inequalities in two-dimensional conformal field theory},  
Rev. Math. Phys. {\bf 17}, 577 (2005), arXiv:math-ph/0412028. 

\bibitem{fgo92} L. H. Ford, P. G. Grove and A. C. Ottewill, {\it Macroscopic detection of negative-energy fluxes},  
Phys. Rev. D {\bf 46}, 4566 (1992).

\bibitem{FR11}  L. H. Ford and T. A. Roman,  {\it Effects of Vacuum Fluctuation Suppression on Atomic Decay Rates},  
Ann. Phys. {\bf 326} 2294 (2011),  arXiv:0907.1638.
 
 \bibitem{DF18} V. A. De Lorenci, and L. H. Ford, {\it Subvacuum effects on light propagation},  arXiv:1804.10132.

\bibitem{BDF14}
C. H. G. Bessa, V. A. De Lorenci, and L. H. Ford, {\it An Analog Model for Light Propagation in Semiclassical Gravity},  
Phys. Rev. D {\bf 90}, 024036 (2014),  arXiv:1402.6285.

\bibitem{BDFR16}
C. H. G. Bessa, V. A. De Lorenci, L. H. Ford and C. C. H. Ribeiro, {\it A Model for Lightcone Fluctuations due to Stress Tensor Fluctuations},  
Phys. Rev. D {\bf 93}, 064067 (2016), arXiv:1602.03857.
  
\bibitem{Colpa1978}
J.~H.~P.~Colpa,  `{\it Diagonalization of the quadratic boson hamiltonian},
Physica {\bf{93A}}, 327 (1978).

\bibitem{Dawson2006}
S. Dawson, {\it Bounds on Negative Energy Densities in Quantum Field Theories in Flat and Curved Space-times},
PhD thesis, University of York, UK, 2006.

 \bibitem{SFF18}  E. D. Schiappacasse, C. J. Fewster and L. H. Ford, {\it Vacuum Quantum Stress Tensor Fluctuations: A Diagonalization Approach},  
 Phys. Rev. D  {\bf 97}, 025013 (2018), 
 arXiv:1711.09477.
 
 \bibitem{FFR12} C. J. Fewster, L. H. Ford and T. A. Roman,  {\it Probability distributions for quantum stress tensors in four dimensions}, 
 Phys. Rev. D {\bf 85}, 125038 (2012), arXiv:1204.3570,
 
 \bibitem{K17} A. Korolov,  {\it  An Investigation of Negative Energy Densities in Quantum Field Theory: Diagonalization Method and Worked Examples},
  undergraduate senior thesis, Tufts University, 2017.
 
 \bibitem{Stoller}  D. Stoler,   {\it Equivalence Classes of Minimum-Uncertainty Packets II}.  Phys. Rev. D {\bf 4}, 1925 (1971).

 \bibitem{Caves} C. M. Caves,   {\it Quantum-mechanical noise in an interferometer},  Phys. Rev. D {\bf 23}, 1693 (1981).
 
 \bibitem{GN}  C.G. Gerry and P.L. Knight, {\it Introductory Quantum Optics}, (Cambridge University Press, Cambridge UK, 2005), Chapter 7.

\bibitem{Slusher} R. E. Slusher, L. W. Hollberg, B. Yurke, J. C. Mertz, and J. F. Valley, {\it Observation of Squeezed States Generated by Four-Wave Mixing in an Optical Cavity},
Phys. Rev. Lett. {\bf 55}, 2409 (1985).

\bibitem{Wu} L-A. Wu, H. J. Kimble, J. L. Hall, and H. Wu,  {\it Generation of Squeezed States by Parametric Down Conversion},
   Phys. Rev. Lett. {\bf 57}, 2520 (1986).

\bibitem{FF15} C. J. Fewster and L. H. Ford, {\it Probability Distributions for Quantum Stress Tensors Measured in a Finite Time Interval},  
Phys. Rev. D {\bf 92}, 105008 (2015), arXiv1508.02359.
 
\end{thebibliography}
\end{document}